# Strongly Ideal Robust Weyl Semimetals in Cubic Symmetry with Spatial-Inversion Breaking

R. Wang,[1,2] J. Z. Zhao,[1,3] Y. J. Jin,[1] W. P. Xu,[1] L.-Y. Gan,[1,4] X. Z. Wu,[1,2] H. Xu[1,*], and S. Y. Tong[1,5]

[1]*Department of Physics, South University of Science and Technology of China, Shenzhen 518055, P. R. China.*
[2]*Institute for Structure and Function & Department of physics, Chongqing University, Chongqing 400044, P. R. China.*
[3]*Dalian Institute of Chemical Physics, Chinese Academy of Sciences, 116023 Dalian, P. R. China*
[4]*Key Laboratory of Advanced Technology of Materials (Ministry of Education), Superconductivity and New Energy R&D Center, Southwest Jiaotong University, Chengdu, 610031 Sichuan, P. R. China and*
[5]*School of Science and Engineering, The Chinese University of Hong Kong (Shenzhen), 518172 Shenzhen, P. R. China.*

**We show that compounds in a family that possess time-reversal symmetry and share a non-centrosymmetric cubic structure with the space group *F-43m* (No. 216) host robust ideal Weyl semi-metal fermions with desirable topologically protected features. The candidates in this family are compounds with different chemical formulas $AB_2$, ABC, $ABC_2$, and ABCD and their Fermi levels are predominantly populated by nontrivial Weyl fermions. Symmetry of the system requires that the Weyl nodes with opposite chirality are well separated in momentum space. Adjacent Weyl points have the same chirality, thus these Weyl nodes would not annihilate each other with respect to lattice perturbations. As Fermi arcs and surface states connect Weyl nodes with opposite chirality, the large separation of the latter in momentum space guarantees the appearance of very long arcs and surface states. This work demonstrates the use of system symmetry by first-principles calculations as a powerful recipe for discovering new Weyl semi-metals with attractive features whose protected fermions may be candidates of many applications.**

Weyl semimetals (WSMs), in which the quasi-particles are described by massless chiral fermions [1], are the subject of intense recent studies. By breaking either time-reversal or spatial-inversion symmetry, Weyl fermions are realized in semi-metallic systems when two linearly dispersing bands cross near the Fermi level [2-9]. Similar to topological insulators (TIs) [10-12] or topological crystalline insulators (TCIs) [13] in which the helical surface states are protected by time-reversal or crystal symmetry, the Weyl nodes with specific chirality (left- or right-handed) are protected by the interplay of symmetry and electronic band topology [14]. The Weyl nodes always appear in pairs of opposite chirality due to the "no-go theorem" [15, 16], and the nodes may be viewed as "magnetic monopoles" or singular points of the Berry curvature in momentum space [2, 3, 14, 17]. The topology of WSMs is characterized by a Chern number $C = \pm 1$ that is related to the chirality or source (sink) of the Berry curvature [5]. As a consequence, an important hallmark of WSMs is the existence of topologically protected Fermi arcs and surface states [2]. These arcs and states terminate at projections of the bulk Weyl nodes having opposite chirality.

Several WSMs have been predicted theoretically and a few observed experimentally [18-24]. As the most interesting properties of WSMs are the Fermi arcs and surface states with protected properties, one would like to find materials in which the Weyl nodes with opposite Chern numbers are well separated in momentum space. This far separation guarantees very long Fermi arcs. Furthermore, the nontrivial properties of WSMs may vanish because closely spaced Weyl nodes with opposite chirality can be annihilated by lattice imperfections. In this paper, we present a class of new materials that may possess such ideal, robust features of WSMs. This class of materials has the non-magnetic space group *F-43m* (No. 216). Compounds belonging to this class can have widely different chemical compositions such as $AB_2$, $ABC$, $ABC_2$, and $ABCD$; where A, B, C, and D represent different elements (e.g., In,

S, Tl, K, Se, Ga, In As, etc.). In the main text, we will use the compound InSTl with the ABC configuration as an example to illustrate the topologically nontrivial features of this class of materials. Results of other compounds in the family: SeGa$_2$, KPTl$_2$, and KInAsTl are included to show spin-orbit-coupling (SOC) effects. Detailed results of the other compounds are given in the Supplementary Information (SI).

The compound InSTl shown in Fig. 1(a) crystallizes in a face-centered-cubic (FCC) lattice with lattice constant *a*, which is 7.228 Å according to our calculation. The crystal structure consists of interpenetrating In, S, and Tl sublattices, where nearest neighbor sublattices are shifted by a structural vector $(a/4, a/4, a/4)$. The In and Tl sublattices form a rocksalt structure, while S and Tl sublattices form a zinc-blende structure. The crystal lacks mirror planes parallel to *x*, *y* and *z* planes and thus, it lacks spatial-inversion symmetry. The FCC BZ and the corresponding (001) surface BZ are shown in Fig. 1(b), with high-symmetry points indicated.

The robust properties of Weyl fermions in these compounds arise from the non-centrosymmetric nature of the non-magnetic *F-43m* group. Time–reversal and crystal symmetries force Weyl nodes on the (001) or equivalent planes with opposite chirality in the absence of SOC to be located on different high symmetry lines in momentum space. The high symmetry lines are ninety degrees apart and well separated in momentum space. Without SOC, the Weyl nodes are doubly (spin) degenerate, occupying the same point on a high symmetry line (e.g., along X-W) in momentum space. When SOC effects are included, the spin degeneracy is lifted and the Weyl nodes occupy adjacent points, either *along* the high symmetry line or *perpendicular* to this line (see below). Because SOC splitting in these compounds is generally much smaller than the separation of the high symmetry lines in momentum space, the Weyl nodes on the (001) or equivalent planes with opposite chirality are guaranteed to be well separated. This in turn generates very long Fermi arcs. In addition, these compounds possess another very useful property: the Fermi level is predominantly populated by nontrivial Weyl fermions. With this feature, contributions of the axial anomaly do not cancel each other by excitations at different nodes and an anomaly-generated current (via the Chiral Magnetic Effect) exists in parallel electric and magnetic fields. This is a technologically useful property given the need of robust (and potentially commercializable) Weyl materials exhibiting axial anomaly.

We perform first-principles calculations based on density functional theory (DFT) [25, 26] as implemented in the Vienna *Ab initio* Simulation Package [27, 28] within the Perdew-Burke-Ernzerhof (PBE) exchange-correlation functional [29, 30]. The core-valence interactions are treated by the projector augmented wave method [31-33] with a cutoff energy of 600 eV. To check whether there are Weyl nodes or not, the dense k-mesh with a 23×23×23 Monkhorst-Pack [34] grid has been employed in self-consistent calculations and band structure calculations throughout the entire BZ. The topological features are further verified using the Heyd-Scuseria-Ernzerhof (HSE) functional [35]. The HSE functional is generally regarded as being more accurate but we find that very similar results are obtained (see Fig. S3 of SI). Thus, in the following, to save computational time, all results use the PBE functional, as they should provide adequate degree of accuracy. Surface band structures and Fermi arcs are obtained using the Green's function method [36] based on a tight binding (TB) Hamiltonian using the maximally localized Wannier functions (MLWFs) [37, 38] that are projected from bulk Bloch wave functions. This approach is generally used because it has been shown to provide adequate accuracy and saves computational time in obtaining surface electronic properties of semi-infinite crystals [5, 6].

The band structures of InSTl along high-symmetry directions of the BZ in the absence of spin-orbital coupling (SOC) are plotted in Fig. 1(c). The bands show that InSTl possesses semi-metallic features in which bands cross each other with linear dispersion along X-W directions. The crossing points are located very close to the Fermi level, at approximately -0.06 eV, forming electron pockets. Considering crystal symmetry, there are a total of 12 such crossing points along X-W directions at the boundaries of the BZ with wave-vectors $k_x = 2\pi/a$, $k_y = 2\pi/a$ and $k_z = 2\pi/a$, respectively. The exact positions of the Weyl nodes may be determined by searching for the "source" or "sink" of the Berry curvature in momentum space [2]. The topological charge or chirality may be obtained from the Chern number $C$ through integral of the Berry curvature across a surface that encloses a Weyl node. As Figs. 2(a) shows, two crossings locate exactly on the $k_x(k_y)$ axis in the $k_z = 2\pi/a$ plane with topological Chern numbers $C=+2$ and $C=-2$, respectively. This indicates that each crossing is a superposition of two Weyl nodes with the same chirality, called a double Weyl node [39].

The effect of SOC is to produce a gap in these crossing points along the high symmetry directions. With SOC, shown in Fig. 1(d), the spin degeneracy along the high symmetry directions is lifted. Each double Weyl node splits into two single Weyl nodes with the same chirality. Band structure calculations throughout the BZ reveal eight single Weyl nodes in the $k_z = 2\pi/a$ plane. Altogether, there are 24 (i.e., 12 pairs of) such single Weyl nodes distributed in the $k_x = 2\pi/a$, $k_y = 2\pi/a$ and $k_z = 2\pi/a$ planes of the FCC BZ [see Fig. 1(b)]. For InSTl, the single Weyl nodes distribute on both sides of the $k_x(k_y)$ axis. Each single Weyl node either has a topological Chern number $C=+1$ or $C=-1$ [see Fig. 2(b)]. The momentum space locations ($\mathbf{k}_{\text{Weyl}}$) of the single Weyl nodes with chirality $C=+1$ are $(2\pi/a, \pm k_1, \pm k_2)$, $(\pm k_2, 2\pi/a, \pm k_1)$, and $(\pm k_1, \pm k_2, 2\pi/a)$, and those with chirality $C=-1$ are $(2\pi/a, \pm k_2, \pm k_1)$, $(\pm k_1, 2\pi/a, \pm k_2)$, and $(\pm k_2, \pm k_1, 2\pi/a)$, respectively, where $k_1$=0.2347 Å$^{-1}$=0.55 $\pi/a$ and $k_2$=0.0956 Å$^{-1}$=0.22 $\pi/a$.

We also present energy dispersions $E(k_x, k_y)$ for InSTl in the $k_z = 2\pi/a$ plane without and with SOC effects in Figs. 3(a) and 3(b), respectively. It can be seen that at each Weyl point, two bands cross with near linear dispersion. Each Dirac cone is anisotropic and the low energy excitations near each Weyl point ($\mathbf{k}_{\text{weyl}}$) are described

by a 2×2 Hamiltonian:

$$H(\mathbf{q}) = \Lambda_0(\mathbf{q})\mathbf{I} + \sum_{i=x,y,z}\Lambda_i(\mathbf{q})\sigma_i \qquad (1)$$

where $\mathbf{q}=\mathbf{k}-\mathbf{k}_{Weyl}$, $\mathbf{I}$ is the unit matrix, and $\sigma_i$ denotes Pauli matrices. The coexistence of time-reversal and crystal symmetry leads to real functions $\Lambda_x(q_x,q_y,0) = \Lambda_y(0,q_y,q_z) = \Lambda_z(q_x,0,q_z)$.

To understand the band topology of symmetry-protected Weyl nodes in the InSTl compound, we plot the orbital-resolved band structures without and with SOC along $W'-X-W$ direction [see Fig. 2(b)] in Figs. 3(c) and 3(d), respectively. The band structures show a band inversion at X, with an inversion gap of approximately 1.5 eV. The valence and conduction orbitals at the band crossings are dominated by S $p_y$ (red) and In&Tl $p_y$ (blue) orbitals, respectively. The effect of SOC splits the spin degeneracy and opens an energy gap of ~0.05 eV in high symmetry X-W directions. Meanwhile, two single Weyl nodes with the same chirality appear along X and W' [Figs. 2 (b) and 3(d)], separated by ~0.19 Å$^{-1}$ in momentum space. The stability of these WSMs may be gauged by the separation in momentum space between two closest Weyl nodes with opposite chirality. In InSTl, the Weyl nodes with opposite chirality are well separated: the separation distance is about 0.20 Å$^{-1}$, which is nearly 23.1% of the reciprocal lattice constant (respectively over two times and four times longer than that found in TaAs [5, 6] and chalcopyrites materials [9]).

Other compounds, such as SeGa$_2$, KPTl$_2$, and KInAsTl in this family are also strongly robust WSMs. Without SOC, the bands cross near the Fermi level along high symmetry (X-W) directions. The crossing points have topological Chern number $C=+2$ or $C=-2$. In the presence of SOC, the double Weyl nodes split into two single Weyl nodes possessing topological charge $C=+1$ or $C=-1$. For the InSTI and KPTl$_2$ compounds, the splitting direction induced by SOC is perpendicular to the X-W direction: $\mathbf{k}_{weyl}$ with chirality $C=+1$ are at $(2\pi/a,\pm k_1,\pm k_2)$, $(\pm k_2,2\pi/a,\pm k_1)$, and $(\pm k_1,\pm k_2,2\pi/a)$, and those with chirality $C=-1$ are at $(2\pi/a,\pm k_2,\pm k_1)$,

$(\pm k_1, 2\pi/a, \pm k_2)$, and $(\pm k_2, \pm k_1, 2\pi/a)$, respectively [Fig. 2(b)]. For the SeGa$_2$ and KInAsTl compounds, the splitting direction is along the X-W direction: $\mathbf{k}_{\text{weyl}}$ with chirality $C=+1$ are at $[2\pi/a, \pm k_1(k_2), 0]$, $[0, 2\pi/a, \pm k_1(k_2)]$, and $[\pm k_1(k_2), 0, 2\pi/a]$, and those with chirality $C=-1$ are at $[2\pi/a, 0, \pm k_1(k_2)]$, $[\pm k_1(k_2), 2\pi/a, 0]$, and $[0, \pm k_1(k_2), 2\pi/a]$, respectively [see Figs. S5 and S6 of SI]. The precise positions of the single Weyl nodes expressed by parameters $k_1$ and $k_2$ for the four compounds are shown in Table 1. Results of phonon dispersion curves indicate that all four compounds are structurally stable (see Fig. S2 of SI).

**Table 1. Lattice constant *a* and positions of Weyl nodes characterized by two parameters $k_1$ and $k_2$ in candidate WSMs. "Perp" and "para" indicate SOC splitting of the Weyl nodes is perpendicular or parallel to the high symmetry line.**

| Compounds | SOC splitting | Chemical formula | $a$ (Å) | $k_1$ (Å$^{-1}$) | $k_2$ (Å$^{-1}$) |
|---|---|---|---|---|---|
| InSTl | (perp) | ABC | 7.228 | 0.2347 | 0.0956 |
| KPTl$_2$ | (perp) | ABC$_2$ | 7.735 | 0.3412 | 0.0203 |
| SeGa$_2$ | (para) | AB$_2$ | 6.683 | 0.4043 | 0.3996 |
| KInAsTl | (para) | ABCD | 7.787 | 0.3668 | 0.3144 |

The well separated Weyl nodes with opposite Chern numbers in non-magnetic *F-43m* group materials leads to very long fermi arcs. To directly illustrate this, we calculate the surface states and Fermi arcs of InSTl using the Green's function method [36] based on the TB Hamiltonian obtained from MLWFs [37, 38]. This is a generally used approach with required accuracy to obtain surface electronic properties of semi-infinite crystals [5, 6]. The calculated local density of states (LDOS) and Fermi surfaces for the semi-infinite (001) surface are shown in Fig. 4. Without SOC, the crossing points along $\overline{\Gamma} - \overline{W}_p$ and $\overline{\Gamma} - \overline{W}_m$ possess Chern numbers +2 and -2, respectively. The surface states connecting two crossing points with opposite Chern numbers are shown in Figs. 4(a), and the corresponding Fermi surface of the (001) surface is shown in Fig. 4(b). There are two Fermi arcs that start and end at projections of two double Weyl nodes with opposite chirality. In the presence of SOC, the double Weyl points split into two single Weyl nodes with Chern number +1 (or -1). Therefore, the surface states and Fermi arcs are also split, as shown in Figs. 4(c) and 4(d), respectively. Here, we only show the projected band structures along $\overline{W}_{p1} - \overline{\Gamma} - \overline{W}_{p2}$. The Fermi arcs connect two single Weyl nodes that are axially symmetric with respect to $\overline{k}_x = -\overline{k}_y$ in the BZ of the (001) surface. The symmetry of the system ensures that the Fermi arcs that cross both the $\overline{k}_x$ and $\overline{k}_y$ axes are very long. The behavior of double vs single Weyl nodes in the compound $KPTl_2$ is the same as that in InSTl. The other compounds in the family, $SeGa_2$ and KInAsTl, whose double Weyl nodes are split by SOC into single Weyl nodes located along the X-W direction, have equally long Fermi arcs. Their long Fermi arcs may be seen in Figs. S5 and S6 of the SI. Such robust features, shared among the four compounds in this family, offer favorable candidates for detection and detailed measurements by experiment in comparison with previously proposed WSMs [5, 6, 8, 9].

In conclusion, we have shown that compounds that possess time-reversal symmetry and share a non-centrosymmetric cubic structure with the space group *F-43m* (No. 216) host robust ideal Weyl semi-metal fermions. The candidates are found to be generalized in compounds with chemical formulas ABC, $AB_2$, $ABC_2$, and ABCD. Representative compounds InSTl, $SeGa_2$, $KPTl_2$, and KInAsTl are chosen to show the topologically nontrivial features. The Fermi level of these compounds is predominantly populated by nontrivial Weyl fermions. The symmetry of the system ensures that Weyl nodes with opposite chirality are well separated in momentum space, resulting in very long Fermi arcs. The robust WSM features, stable against annihilation by lattice perturbations, are promising candidates for being exploited in applications utilizing their nontrivial properties. This work demonstrates that system symmetry is a powerful recipe for discovering desirable WSM features for future research.

This work is supported by the National Natural Science Foundation of China (NSFC, Grant Nos. 11674148, 11304403, 11334003 and 11404159) and the Basic Research Program of Science, Technology and Innovation Commission of Shenzhen Municipality (Grant No. JCYJ20160531190054083).

**Author Contributions**

**Equal Contributions:**

R. Wang and J. Z. Zhao contributed equally to this work.

**Corresponding author:**

*Email: xuh@sustc.edu.cn (H. X.)

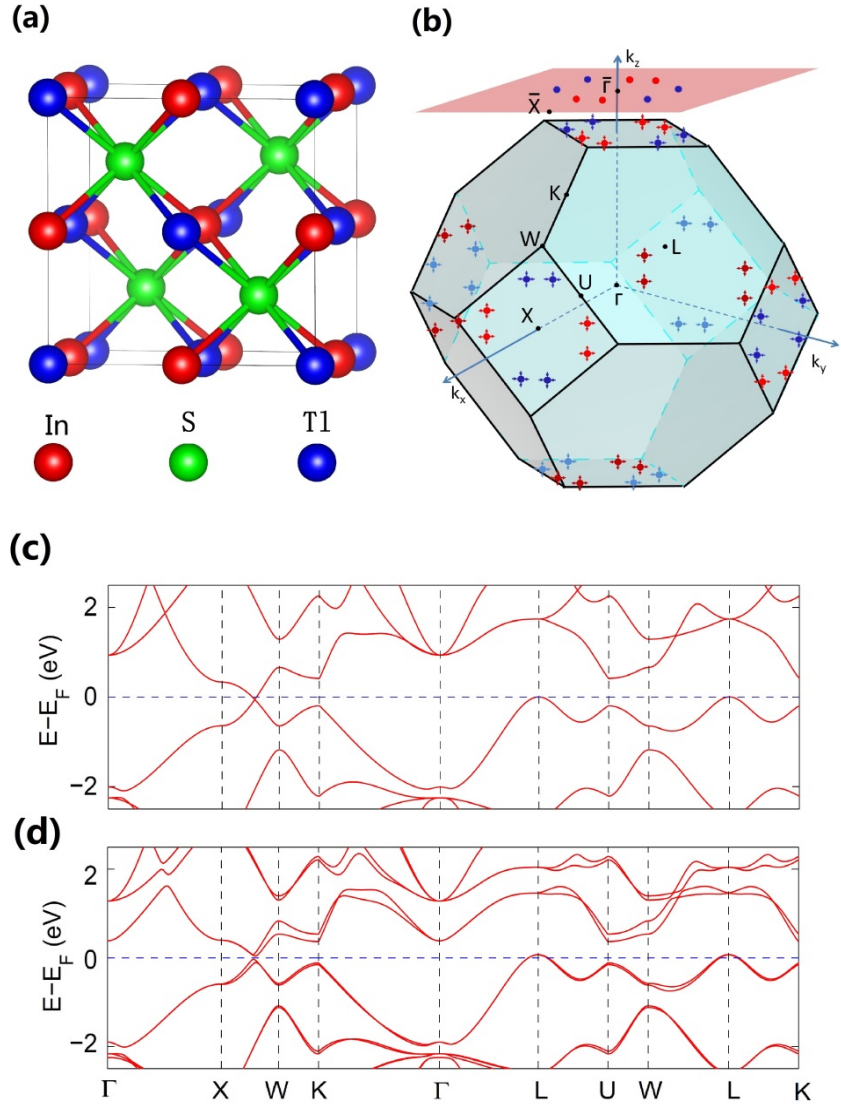

FIG. 1. (a) Crystal structure of the InSTl compound with space group *F-43m* (No. 216). (b) The FCC BZ (blue) and the corresponding (001) surface BZ (pink). The crosses represent single Weyl nodes in the presence of SOC, red (C=+1) and blue (C=-1) and dots are the Weyl nodes projected onto the surface BZ. (c) Band structures of InSTl without spin-orbital coupling (SOC), and (d) with SOC.

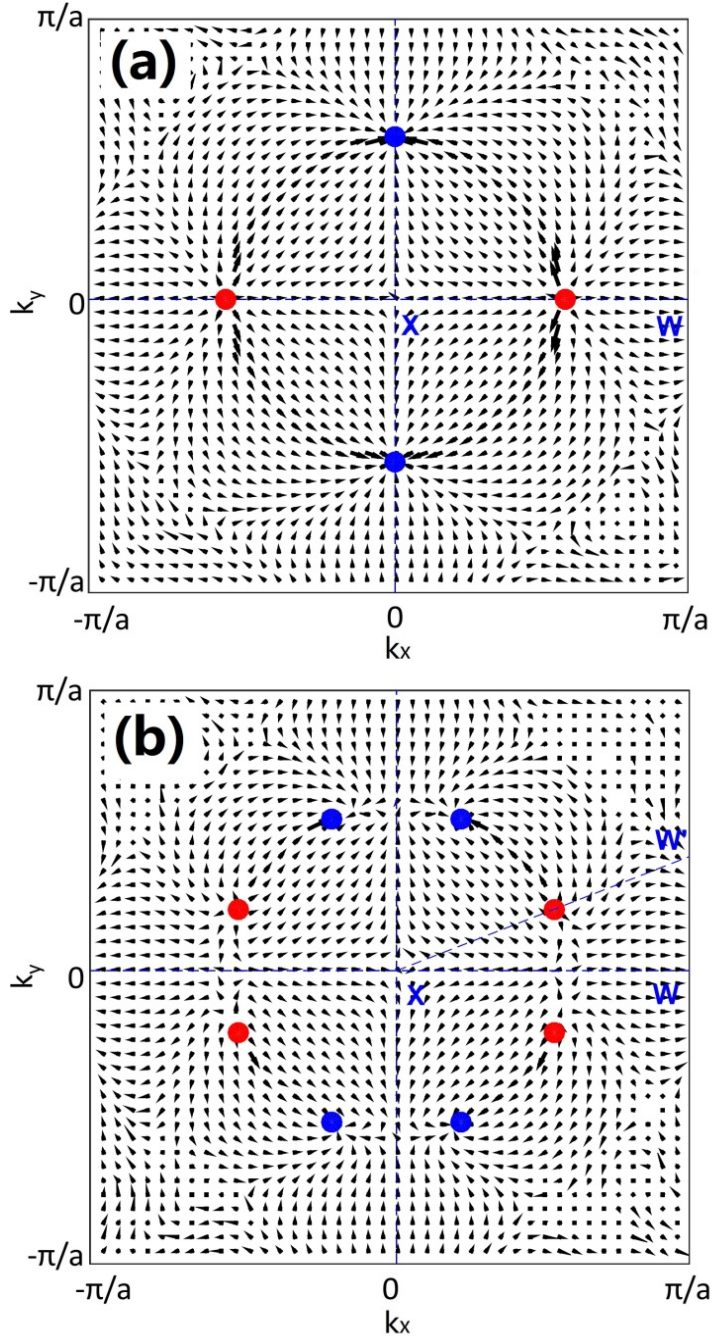

FIG. 2. The Berry curvature of InSTl (a) without SOC and (b) with SOC in the $k_z = 2\pi/a$ plane, where the red and blue dots denote Weyl nodes with positive or negative Chern numbers, respectively. Without SOC, the Weyl nodes with Chern number $C=+2$ ($C=-2$) locate on the $k_x$ ($k_y$) axis, as shown in (a). With SOC, the doubly degenerate Weyl nodes split and single Weyl nodes with Chern number $C=+1$ ($C=-1$), respectively distribute on both sides of the $k_x$ ($k_y$) axis [as shown in (b)].

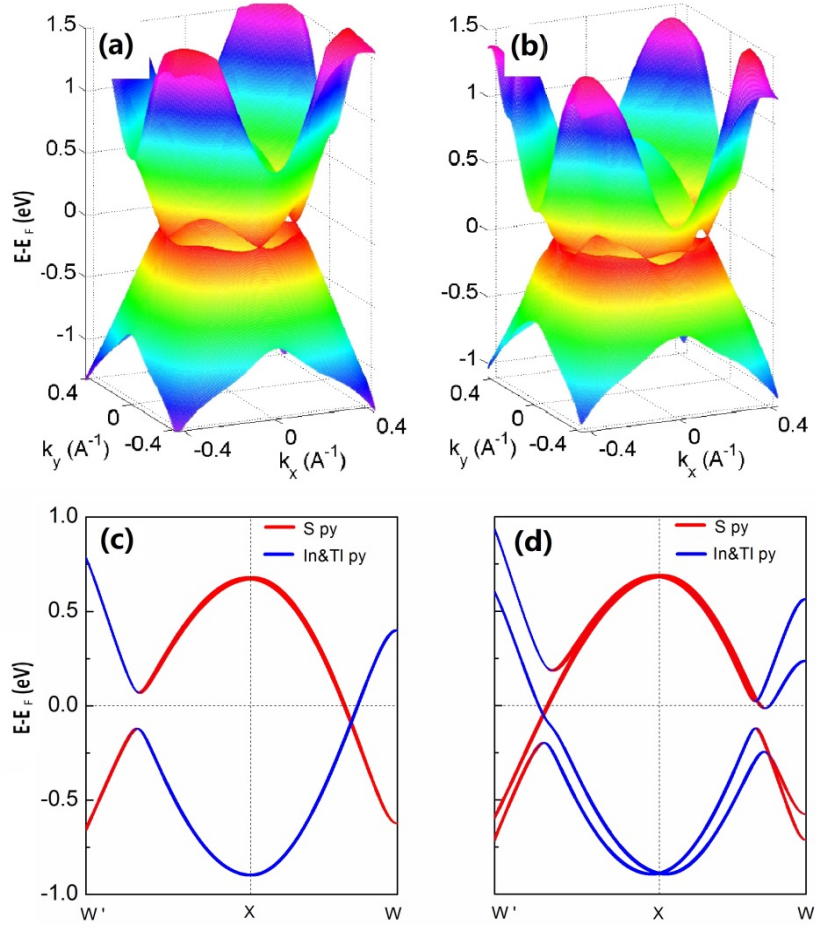

FIG. 3. Energy dispersion E ($k_x$, $k_y$) in the $k_z = 2\pi/a$ plane without (a) and with (b) SOC effect are shown. (c) Orbital-resolved band structures of InSTl without SOC along W'-X-W directions are shown. The component of S$p_y$ (In&Tl $p_y$) orbitals is proportional to the width of the red (blue) curve. (d) Orbital-resolved band structures of InSTl with SOC along W'-X-W directions are shown. An energy gap opens between X and W and a Weyl node (crossing) appears between X and W'.

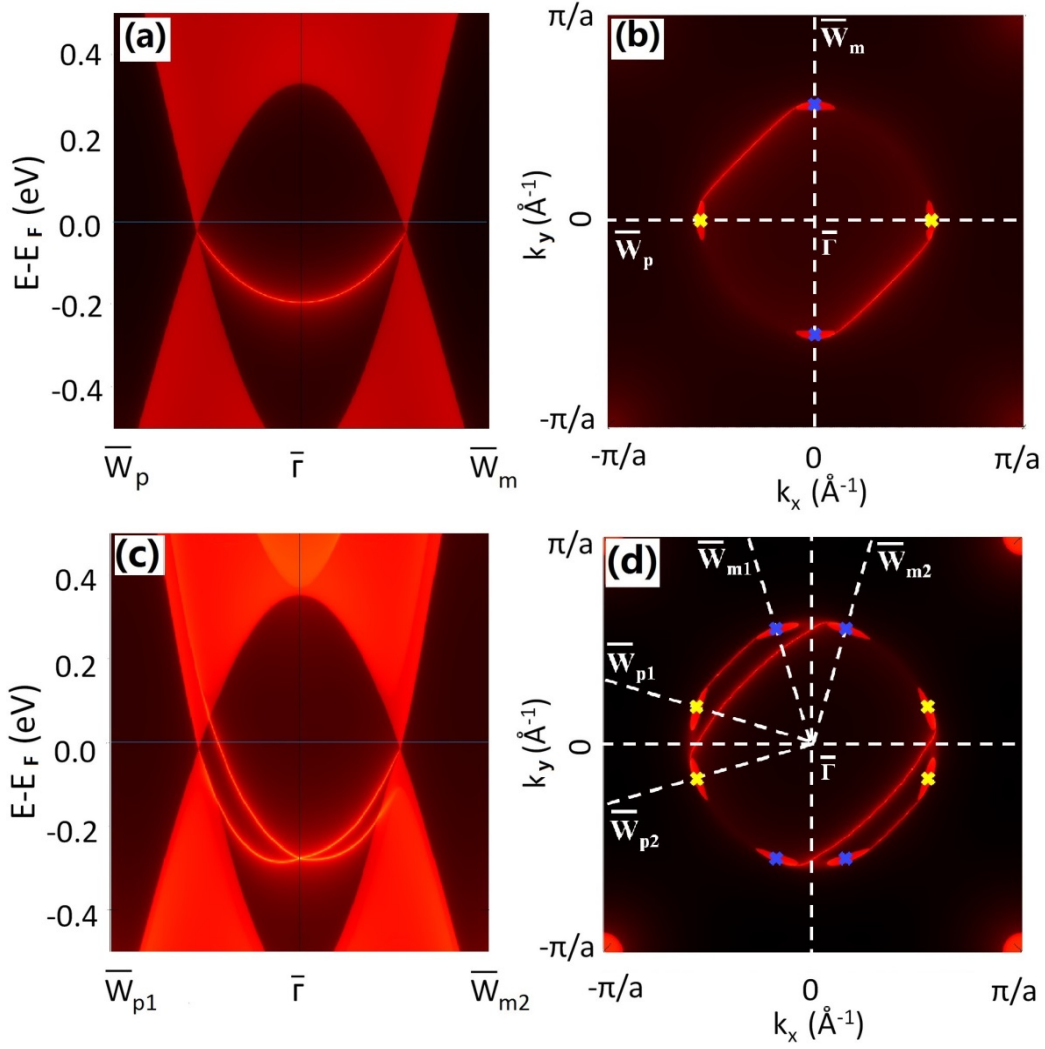

FIG. 4. Surface states and Fermi arcs projected onto the (001) surface of InSTl. (a) Local density of states (LDOS) and (b) the Fermi surface without SOC. The double Weyl nodes projected along $\bar{\Gamma}-\bar{W}_p$ and $\bar{\Gamma}-\bar{W}_m$ have opposite chirality. (c) LDOS and (d) Fermi surface with SOC. The double Weyl nodes are split into two single Weyl nodes projected along $\bar{\Gamma}-\bar{W}_{p1}$ and $\bar{\Gamma}-\bar{W}_{p2}$ (or $\bar{\Gamma}-\bar{W}_{m1}$ and $\bar{\Gamma}-\bar{W}_{m2}$). In (a) and (c), red regions represent the bulk bands, and red lines represent surface states. In (b) and (d), red lines denote Fermi arcs connecting two projected Weyl points with positive (yellow) and negative (blue) chirality, respectively.

# The Supplemental Information for
# Strongly Robust Ideal Weyl Semimetals in Cubic Symmetry with Spatial-Inversion Breaking

R. Wang,[1,2] J. Z. Zhao,[1,3] Y. J. Jin,[1] W. P. Xu,[1] L.-Y. Gan,[1,4] X. Z. Wu,[1,2] Hu Xu[1,*], and S. Y. Tong[1,5]

[1]*Department of Physics, South University of Science and Technology of China, Shenzhen 518055, P. R. China.*
[2]*Institute for Structure and Function & Department of physics, Chongqing University, Chongqing 400044, P. R. China.*
[3]*Dalian Institute of Chemical Physics, Chinese Academy of Sciences, 116023 Dalian, P. R. China*
[4]*Key Laboratory of Advanced Technology of Materials (Ministry of Education), Superconductivity and New Energy R&D Center, Southwest Jiaotong University, Chengdu, 610031 Sichuan, P. R. China and*
[5]*School of Science and Engineering, The Chinese University of Hong Kong (Shenzhen), 518172 Shenzhen, P. R. China.*

*Email: xuh@sustc.edu.cn

## 1. The crystal structures and phonon dispersions

The compounds $SeGa_2$, $KPTl_2$, and $KInAsTl$ belong to the chemical formulas $AB_2$, $ABC_2$, and $ABCD$, respectively. The structures of these three compounds are shown in Fig. S1. The crystal structure with space group *F-43m* consists of four possible interpenetrating sublattices, positioned at (0, 0, 0), (0.25, 0.25, 0.25), (0.5, 0.5, 0.5), and (0.75, 0.75, 0.75), respectively. For $SeGa_2$, Se atom is positioned at (0, 0, 0), two Ga atoms are positioned at (0.25, 0.25, 0.25) and (0.50, 0.50, 0.50), respectively. For $KPTl_2$, K atom is positioned at (0, 0, 0), two Tl atoms are positioned at (0.25, 0.25, 0.25) and (0.50, 0.50, 0.50), and P atom is positioned at (0.75, 0.75, 0.75). For $KInAsTl$, K, In, Tl, and As atoms are positioned at (0, 0, 0), (0.25, 0.25, 0.25), (0.50, 0.50, 0.50), and (0.75, 0.75, 0.75), respectively.

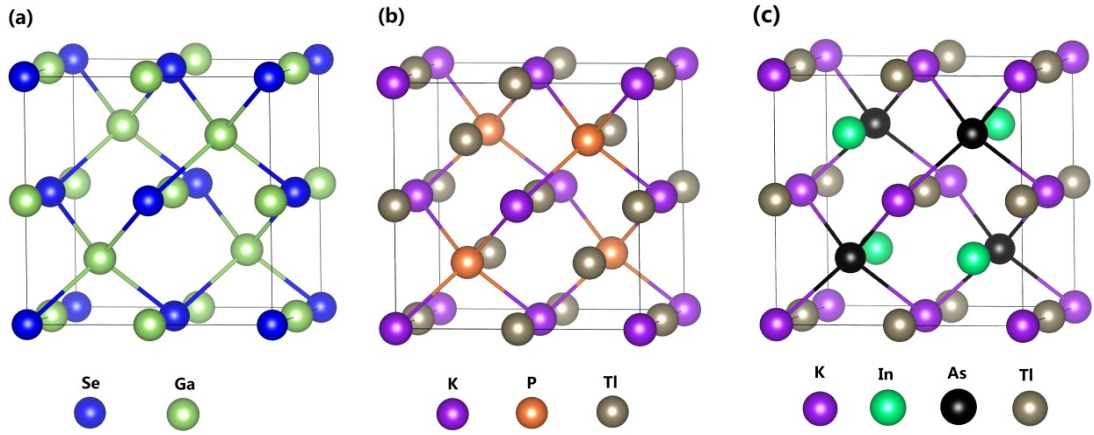

**FIG. S1.** The crystal structures of (a) SeGa$_2$, (b) KPTl$_2$, and (c) KInAsTl.

The phonon spectrum is one useful way to investigate the stability and structural rigidity. The phonon dispersions of InSTl, SeGa$_2$, KPTl$_2$, and KInAsTl are shown Fig. S2. The method of force constants has been used to calculate the phonon frequencies as implemented in PHONOPY package [1-3]. We find that there is the absence of any imaginary frequencies over the entire BZ, suggesting that these candidates are dynamical stability.

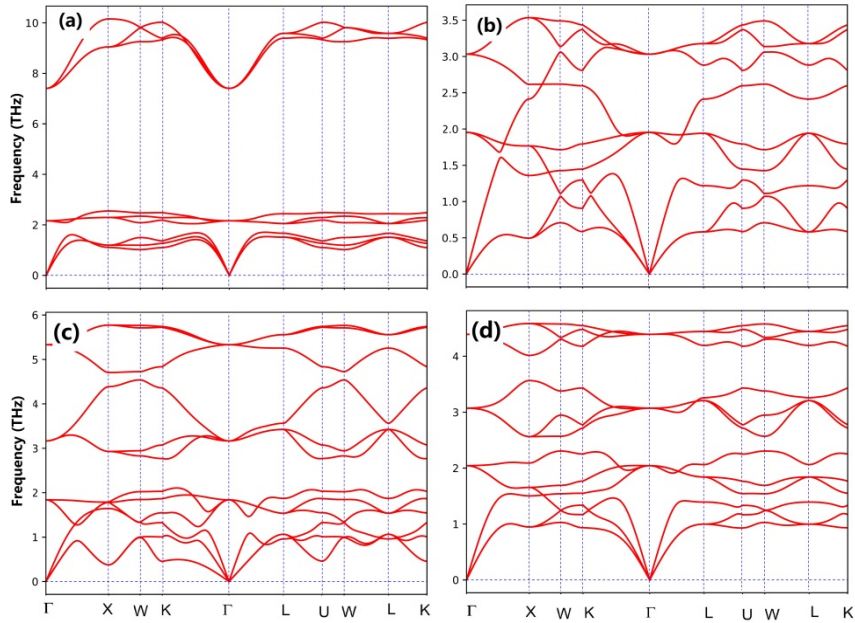

**FIG. S2.** The calculated phonon band structures of (a) InSTl, (b) SeGa$_2$, (c) KPTl$_2$, and (d) KInAsTl.

## 2. The HSE band structures of InSTl compounds

We further use the Heyd-Scuseria-Ernzerhof (HSE) functional [4] to confirm the topological features. Fig. S3 shows the band structures of InSTl in both PBE and HSE calculations. It is found that HSE calculations essentially give the very similar results in comparison with PBE calculations.

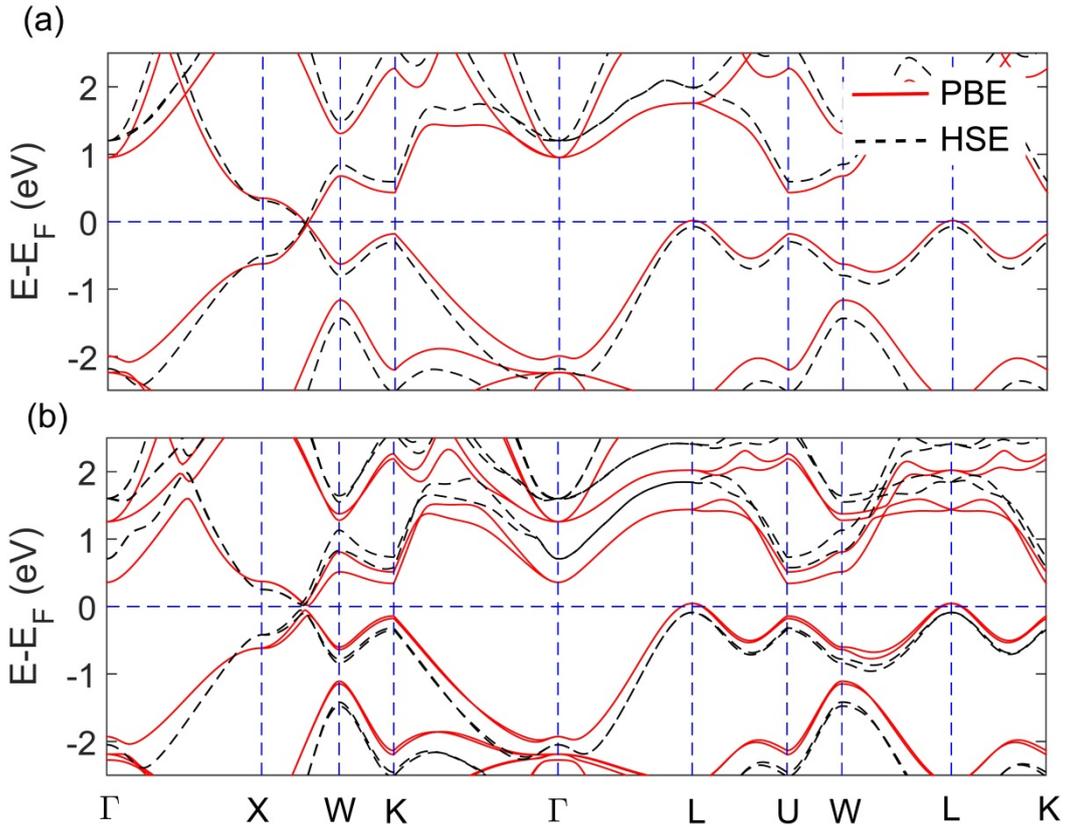

FIG. S3. The PBE and HSE band structures of InSTl (a) without SOC and (b) with SOC. The solid (red) and dashed (black) lines represent the results of PBE and HSE, respectively.

## 3. The band structures and Fermi arcs of $KPTl_2$

The band structures of $KPTl_2$ without and with SOC are shown in Figs. S4(a) and S4(b). The splitting direction induced by SOC is perpendicular to X-W direction. Two neighboring Weyl points with same chirality split by ~0.04 Å$^{-1}$ in momentum space, while the separation distance for the Weyl points with opposite chirality is about 0.45 Å$^{-1}$. The Fermi surface projected on the (001) surface is shown Fig. S4(c).

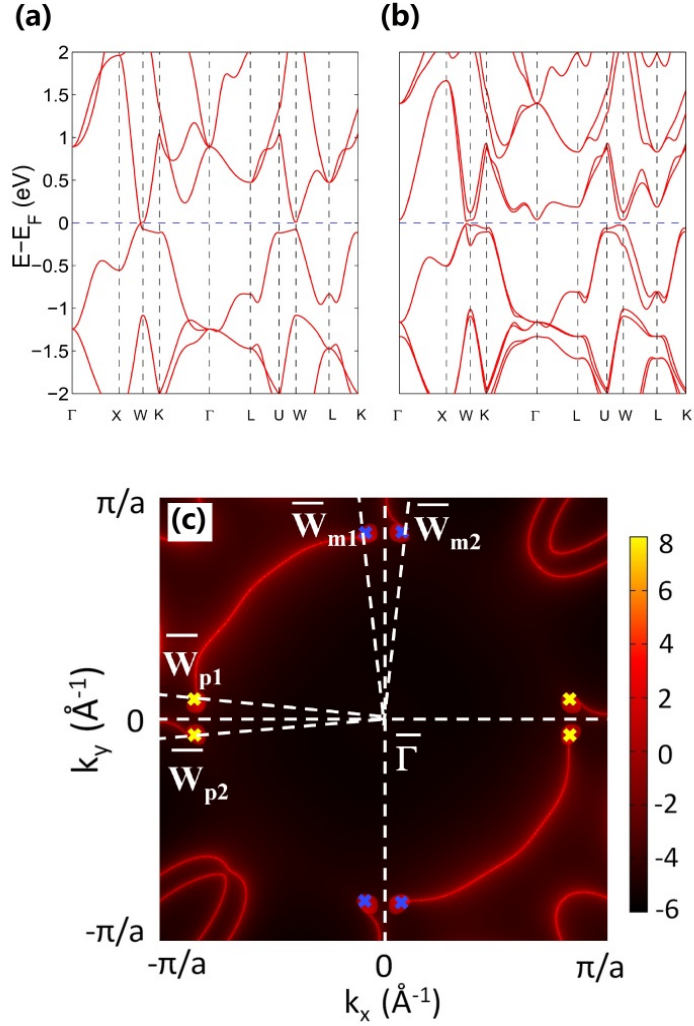

**FIG. S4. Band structures of KPTl2 without (a) and with (b) SOC. (c) Fermi surface projected on (001) surface with SOC. The double Weyl points are split into two single Weyl points projected along $\overline{\Gamma}-\overline{W}_{p1}$ and $\overline{\Gamma}-\overline{W}_{p2}$ (or $\overline{\Gamma}-\overline{W}_{m1}$ and $\overline{\Gamma}-\overline{W}_{m2}$). Red regions represent the bulk bands, and single red lines represent surface states. The red lines denote Fermi arcs connect two projected Weyl points with positive (yellow) and negative (blue) chirality, respectively.**

## 4. The band structures and Fermi arcs of SeGa$_2$

The band structures of SeGa$_2$ without and with SOC are shown in Figs. S5(a) and S5(b). The splitting direction induced by SOC is along X-W direction. Two neighboring Weyl points with same chirality split by ~0.005 Å$^{-1}$ in momentum space, while the separation distance for the Weyl points with opposite chirality is about 0.57 Å$^{-1}$. The fermi surface projected on the (001) surface is shown Fig. S5(c).

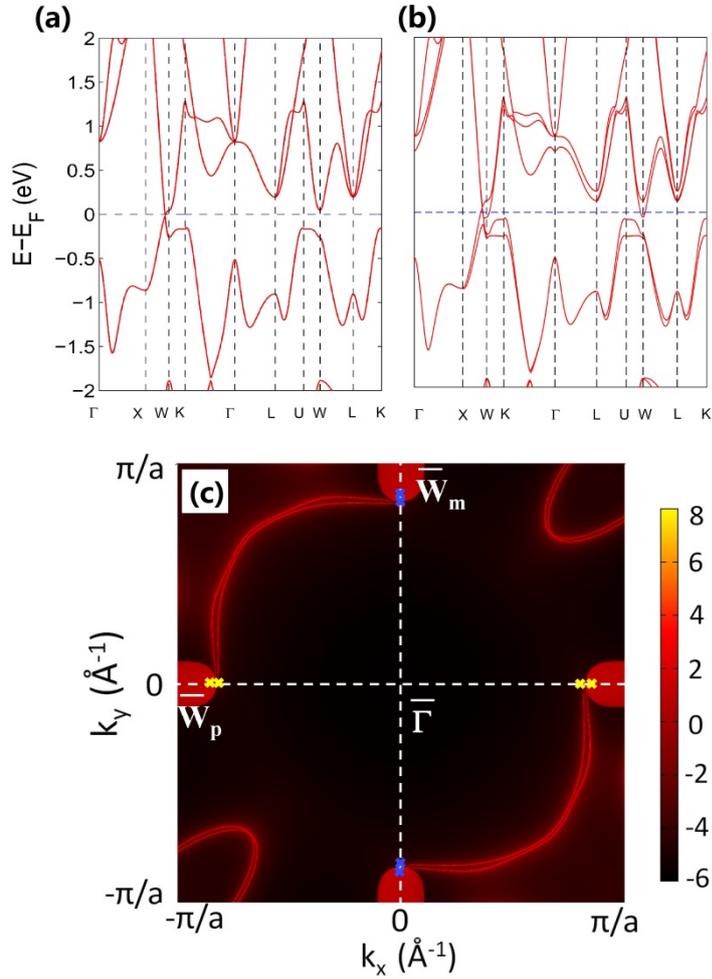

**FIG. S5. Band structures of SeGa$_2$ without (a) and with (b) SOC. (c) Fermi surface projected on (001) surface with SOC. The double Weyl points are split into two single Weyl points projected along $\overline{\Gamma}-\overline{W}_p$ and $\overline{\Gamma}-\overline{W}_m$. Red regions represent the bulk bands, and single red lines represent surface states. The red lines denote Fermi arcs connect two projected Weyl points with positive (yellow)**

and negative (blue) chirality, respectively.

## 5. The band structures and Fermi arcs of KInAsTl

The band structures of KInAsTl without and with SOC are shown in Figs. S6(a) and S6(b). The splitting direction induced by SOC is along X-W direction. Two neighboring Weyl points with same chirality split by ~0.05 Å$^{-1}$ in momentum space, while the separation distance for the Weyl points with opposite chirality is about 0.44 Å$^{-1}$. The fermi surface projected on the (001) surface is shown Fig. S6(c).

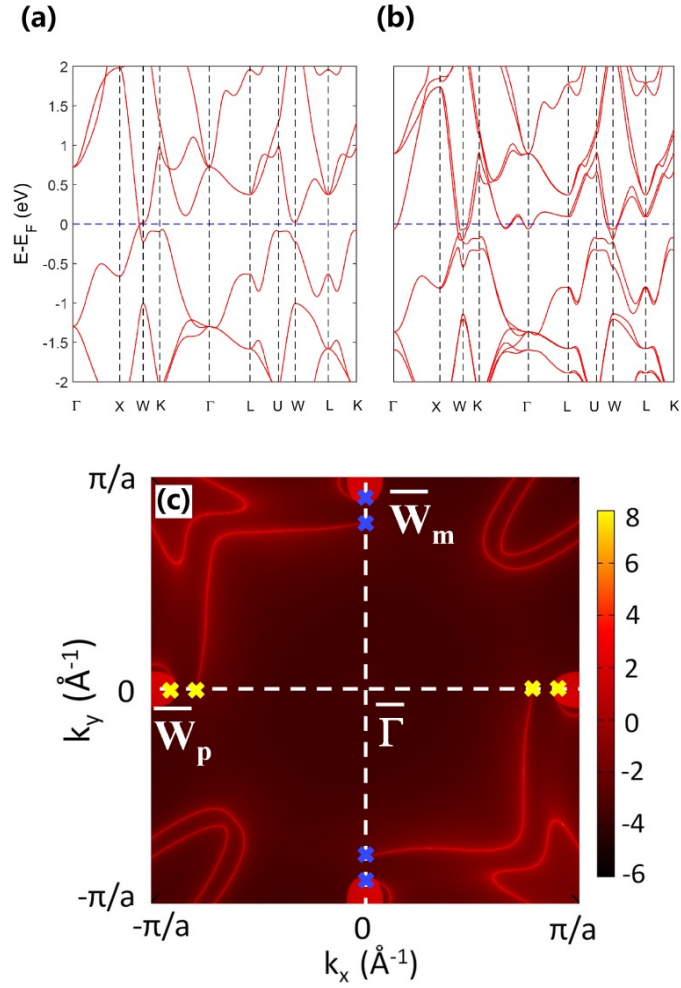

**FIG. S6. Band structures of KInAsTl without (a) and with (b) SOC. (c) Fermi surface projected on (001) surface with SOC. The double Weyl points are split into two single Weyl points projected along $\overline{\Gamma}-\overline{W}_p$ and $\overline{\Gamma}-\overline{W}_m$. Red regions represent the bulk bands, and single red lines represent surface states. The red**

**lines denote Fermi arcs connect two projected Weyl points with positive (yellow) and negative (blue) chirality, respectively.**